\documentstyle[twoside,fleqn,espcrc2,epsf]{article}

\title{Remarks on the realization of the Atiyah-Singer index theorem 
in lattice gauge theory\thanks{Supported by Fonds zur F\"orderung 
der Wissen\-schaft\-li\-ch\-en Forschung in \"Osterreich, Projects P11502-PHY 
and J01185-PHY.}}

\author{C.~R. Gattringer\address{Department of Physics and Astronomy, 
University of British Columbia, Vancouver B.C., Canada}%
\thanks{Speaker at the conference.},
I. Hip\address{Institut f\"ur Theoretische Physik,
Universit\"at Graz, A-8010 Graz, Austria}%
and C.~B. Lang$^{\rm b}$
}
       
\begin{document}

\begin{abstract}
We discuss the interplay between topologically non-trivial gauge field configurations and the spectrum of the Wilson-Dirac operator 
in lattice gauge theory. Our analysis is based on analytic arguments 
and numerical results from a lattice simulation of QED$_2$. 
\end{abstract}

\maketitle

\section{MOTIVATION}

It has been known \cite{JaRe76}
for more than two decades that topologically non-trivial
configurations play an essential role for a proper understanding of QCD. 
However, it is not straightforward to go beyond a semiclassical analysis
when implementing topological ideas in the continuum path integral. 
There are at least two reasons: Firstly
the continuum path integral (if it can be constructed at all) has support 
only on configurations without the necessary smoothness condition for 
a definition of the topological charge - in general the 
gauge fields contributing to the continuum path integral are simply too 
rough to be classified with respect to their topological charge.
Secondly one would also like to use the 
Atiyah-Singer index theorem \cite{AtSi71} in the fully quantized theory. 
It relates the topological charge of a 
differentiable background configuration to the zero modes of the Dirac 
operator. Again the lack of smoothness of the gauge fields contributing 
to the path integral does not allow the application of the index theorem 
beyond a semiclassical analysis.

On the lattice the situation is different. There are several definitions 
of the topological charge on the lattice which allow to classify (almost)
all gauge field configurations contributing
in the continuum limit. In contrary to the continuum
a decomposition of the path integral into 
topological sectors is possible. Concerning the Atiyah-Singer
index theorem, the situation is more difficult. 
There is no analytic result for an index theorem on the lattice, but it 
is generally believed that an equivalent result should be manifest 
on the lattice in a probabilistic way.

Although several aspects of the interplay bet\-ween topologically non-trivial
configurations and the spectrum of the Dirac operator on the lattice
have been analyzed \cite{KaSeSt86}, 
due to the immense computational cost no complete
study in fully quantized QCD$_4$ has been accomplished. The 
numerical data which
we present in this contribution come from a study 
\cite{GaHiLa97} in lattice QED$_2$. 
This model has similar structural features as 
QCD$_4$ (topologically non-trivial configurations, existence of the anomaly,
U(1)-problem etc.) but is computationally much less demanding.
This allows for a proper understanding of the 
role of topology in that simple model, and indicates what to expect for 
a lattice study of QCD$_4$.

\section{SYMMETRIES OF THE WILSON-DIRAC OPERATOR}

Before numerically analyzing the spectrum of the Wilson-Dirac 
operator $M$ it is helpful to reassess the symmetry properties of the
operator (fermion matrix) and thus of the spectrum. We denote the
fermion matrix as $M = 1 - \kappa Q$, where $\kappa$ is the hopping 
parameter and $Q$ the hopping matrix 
\begin{equation}
Q (x,y) = \sum_{\nu = \pm 1}^{\pm D} 
(1+\gamma_\nu) U_\nu (x - \hat{\nu}) \; \delta_{x-\hat{\nu},y} \; ,
\label{Qferm} 
\end{equation}
where we use the notation $\gamma_{-\nu} = - \gamma_\nu$ and 
$U_{-\nu}(x + \hat{\nu}) = U_\nu (x)^\dagger$. $D$ is the number of 
dimensions (2 or 4) and $U_\nu(x)$ are elements of the gauge group
(U(1) for D = 2 and SU(N) for D = 4). We work on a finite 
$L^{D-1} \times T$ lattice with mixed periodic boundary conditions. 
The existence of the following similarity transformations is well known \cite{KaSeSt86}
\begin{equation}
\Gamma_5\, Q \,\Gamma_5 = Q^\dagger \; , \; \Xi \,Q \,\Xi  =  -Q \;,
\label{simtra}
\end{equation}
where the latter only holds for even $L$ and $T$. The matrices 
$\Gamma_5 = \gamma_5\,\delta_{x,y}$ and $\Xi = (-1)^{x_1 + .. + x_D}\,
\delta_{x,y}$ are both unitary and hermitian. This implies that the 
spectrum of $Q$ is symmetric with respect to reflection on both the
real and the imaginary axis. Thus the eigenvalues of the hopping matrix
on a lattice with even $L$ and $T$ come in complex quadruples or in 
real pairs.

The special role of the {\it real} eigenvalues $\lambda$
of the hopping matrix (and thus the 
full fermion matrix $M = 1 - \kappa \,Q$) can be appreciated only by 
understanding the chiral 
properties of their eigenvectors. 
Denote by $v_\lambda, v_\mu$ the eigenvectors of $Q$ with 
eigenvalues $\lambda, \mu$. Using the symmetries (\ref{simtra}) it
can be shown \cite{GaHiLa97} that 
\begin{equation}
{v_\mu}^\dagger \, \Gamma_5 \, v_\lambda \; \neq \; 0 \; ,
\end{equation}
only for $\mu = \overline{\lambda}$. In particular for the diagonal 
elements (this result can already be found in \cite{NaNe95})
\begin{equation}
{v_\lambda}^\dagger \; \Gamma_5 \; v_\lambda \; \neq \; 0 \; ,
\end{equation} only for {\it real} eigenvalues $\lambda$. This 
result has to be compared to the continuum result 
($\psi$ denotes some eigenstate of the continuum Dirac operator)
\begin{equation}
(\psi, \gamma_5 \psi) \; \neq \; 0 \; ,
\end{equation} 
only if $\psi$ is a {\it zero mode}. This indicates that only the 
eigenvectors of the fermion matrix corresponding to real eigenvalues
can play the role of the zero modes of the continuum.

At this point we would like to remark that for an analysis of the
realizations of index theorems on the lattice the Wilson form of 
the lattice Dirac operator is superior to the staggered version. The
latter is antihermitian and thus has only purely complex eigenvalues
lacking a straightforward criterion for identifying zero modes.
Furthermore it is also not clear how to relate
the (real) eigenvalues of the hermitian 
operator $\Gamma_5 Q$ to the spectrum of $Q$ with
its more sophisticated features.   

\section{NUMERICAL RESULTS IN QED$_2$}

Having established the real eigenvalues of the fermion matrix as the trace
of gauge field configurations with non-trivial topology one can simplify a 
numerical analysis by concentrating on the real eigenvalues alone. Here 
we discuss results from a numerical simulation of QED$_2$ on the lattice 
\cite{GaHiLa97}. In this model the evaluation of the geometric definition 
\cite{Lu82} of the topological charge $\nu[U]$ is straightforward,
giving
\begin{equation}
\nu[U]  =  \frac{1}{2\pi} \sum_{x \in \Lambda} \theta_P(x) 
\; \; \in \; \mbox{Z\hspace{-1.3mm}Z} \; .
\label{toplat}
\end{equation}
The plaquette angle $\theta_P(x)$ is introduced as $\theta_P(x) =  \mbox{Im}\, \ln U_P(x)$ where $U_P$ are the ordered products of the 
link variables $U_\nu$ around the plaquette. In order to understand an
eventual realization of the index theorem on the lattice one would like
to establish a relation between the number of real eigenvalues (compare the
discussion above) of the fermion matrix and the functional $\nu[U]$. 

After inspecting spectra for several 
configurations one can conjecture the rule
\begin{equation}
\#\; \mbox{of~real~eigenvalues}  = 4\, |\nu| \; .
\label{nre}
\end{equation}
It has to be remarked, that this is not a result in a mathematically strict 
sense, since isolated gauge field configurations which violate (\ref{nre})
can be constructed. However, it can be tested numerically whether (\ref{nre})
holds for a considerable portion of the gauge field configurations of
the lattice path integral. In order to analyze this question we performed
a simulation of lattice QED$_2$ with two flavors of dynamical fermions 
at several values of $\beta$ (inverse coupling squared)
and $L (T=L)$. The value of $\kappa$ was always chosen
close to the critical $\kappa$ for the respective values of $\beta$ and $L$.
For each Monte Carlo configuration the topological charge (\ref{toplat})
was evaluated. We define $p(\beta)$ to be the probability of
finding (\ref{nre}) correct for a given $\beta$ (and $L$).

\begin{figure}[htb]
\vspace{-9pt}
\epsfysize=2.3in
\epsfbox[25 165 395 465] {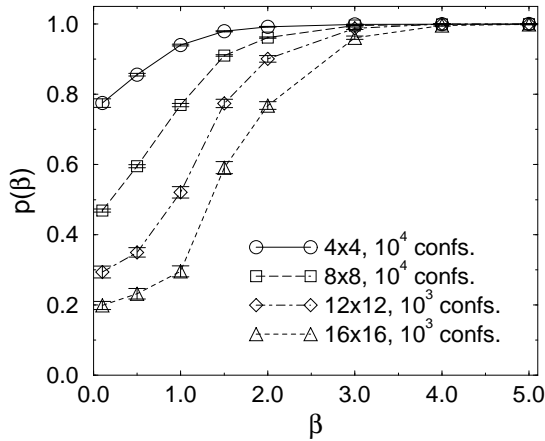}
\vspace{-9mm}
\caption{Probability $p(\beta)$ of finding (\protect{\ref{nre}}) 
correct as a function 
of $\beta$. We show our results for lattice sizes $L=4, 8, 12$ and 16. 
The symbols are connected to guide the eye. 
\label{okvsbeta}}
\vspace{-7mm}
\end{figure}

It is obvious from fig. 1 that already for moderately high values of $\beta$ the 
configurations obeying (\ref{nre}) entirely dominate the lattice path
integral. This establishes, that with the proper interpretation of 
the real eigenvalues and their eigenvectors as the trace of configurations
with non-trivial topology, the index theorem holds in a probabilistic sense.
In particular in the continuum limit $\beta \rightarrow \infty$ (\ref{nre})
holds with probability 1. 

At this point we should  point out the particular features of
QED$_2$ as opposed to QCD$_4$. For QED$_2$ it is known \cite{Ki77}, that in 
addition to the Atiyah Singer index theorem there holds the so-called
{\it vanishing theorem}, which states that there are either only left-handed or 
only right-handed zero modes. Thus the result (\ref{nre}) is exactly what one
expects from the index and vanishing theorems for smooth continuum 
configurations (the factor 4 comes from the extra real eigenvalues 
caused by the doublers). In QCD there is no vanishing theorem. Thus one 
expects the {\it difference} of the numbers of 
left- and right-handed real eigenvalues 
to be related to the topological charge. 

How can one utilize an 
understanding of the realization of the index theorem on the lattice? 
In general all {\it semiclassical} continuum
arguments involving the index theorem can be extended to the {\it fully 
quantized} lattice model. In \cite{GaHiLa97} it was e.g.~demonstrated 
how the dependence of the pseudoscalar density on the topological
charge can be understood using the index theorem. In addition  
one can use the results for the spectrum to analyze problems that are 
specific for the lattice approach. In particular in \cite{GaHiLa97}
we investigated the behaviour of the sign of the fermion determinant,
establishing that it can become negative only for configurations with
non-trivial topology. 

We believe, that extending the numerical part of our analysis to 
QCD$_4$ will give interesting insight on the role of topologically
non-trivial gauge field configurations in the fully quantized model.

\end{document}